\documentclass[a4paper]{revtex4}
\usepackage{graphicx}
\usepackage{fancyhdr}
\usepackage{amsmath}
\usepackage{amssymb} 
\usepackage{amsfonts}
\newcommand{\Z}[1]{{\mathbb Z}_#1}

\usepackage{color}

\begin{document}

\title{Higgs as a probe of supersymmetric grand unification with the
	Hosotani mechanism}

%

\author{T. Yamashita}
\affiliation{Department of Physics,
    Aichi Medical University, Nagakute 480-1195, Japan}

\begin{abstract}
The supersymmetric grand unified theory where the 
$SU(5)$ gauge symmetry is broken by the Hosotani mechanism 
provides a natural solution to the so-called doublet-triplet 
splitting problem.
At the same time, this model derives a general and distinctive
prediction that is testable at TeV scale collider experiments.
To be more concrete, adjoint chiral supermultiplets with masses 
around TeV scale appear.  
Since these additional fields originate from a
higher-dimensional gauge supermultiplet, our model is highly predictive.
We 
study especially the Higgs sector and show 
that our model is discriminative from the others 
by precision measurements of the couplings and masses.
Namely, 
we may get a hint of the breaking mechanism 
of the grand unification at future collider experiments. 
\end{abstract}

\maketitle

\thispagestyle{fancy}


\section{Introduction}
By the discovery of the standard model (SM) like Higgs boson whose 
mass is around 125 GeV, as
reported in 2012 by the ATLAS and CMS collaborations of the CERN Large
Hadron Collider (LHC) \cite{LHC}, the
SM has been established at least as a low energy effective theory. 
It is, however, not the end of the story, as the SM can not explain 
the neutrino
oscillations, existence of dark matter and baryon asymmetry of the
universe.  
In addition, there are several unsatisfactory points from the 
theoretical viewpoint, such as the so-called hierarchy problem and 
charge quantization problem. 
%
These two theoretical problems motivate the supersymmetry (SUSY) 
and the grand unified theory (GUT) \cite{GUT,SUSY-GUT}. 
In particular, 
when these two ideas are assumed simultaneously, the three running 
gauge couplings meet with each others at a very high scale around 
$10^{16}$ GeV, called the GUT scale, in the minimal model. 
%
Thus,
the combination, the SUSY-GUT, is worth examining seriously 
as the physics beyond the SM. 
Then, the so-called doublet-triplet (DT) splitting problem is the 
biggest one to be solved. 
In the minimal model, the mass splitting is realized by fine-tuning, 
but it dismisses the above motivation of the SUSY. 
Beyond the minimal model, many ideas to solve the DT splitting problem 
have been proposed
\cite{DTS}.  
In extended SUSY-GUT models, however, the successful gauge coupling 
unification (GCU) is spoiled in many cases and the GCU becomes a 
constraint instead of a prediction.
In addition, there is an unfavorable point that the typical scale of 
the SUSY-GUTs is too high to address directly at any practical collider 
experiments. 

In these respects, especially the latter one, the model proposed in 
Ref.~\cite{gGHU-DTS} by supersymmetrizing the grand gauge-Higgs 
unification (gGHU)~\cite{gGHU} is attractive. 
The gGHU is a kind of GUTs where the $SU(5)$ gauge symmetry is 
broken by the so-called Hosotani mechanism~\cite{hosotani}.
In the SUSY version of the gGHU (SGGHU), the DT splitting problem is 
naturally solved and, as a by-product, existence of chiral adjoint 
supermultiplets as light as the SUSY-breaking scale is generally 
predicted. 
These additional fields may be tested at future collider 
experiments, 
if the SUSY-breaking scale is around 
TeV scale which is required to solve the hierarchy problem. 
In Ref.~\cite{gGHU-pheno}, we study the phenomenology of the SGGHU, 
especially focusing on its Higgs sector. 


In this article, we review our model in Sec.~\ref{sec:gGHU} and briefly 
describe some features of its low energy effective theory in 
Sec.~\ref{sec:LEET}. 
Then, we focus on its Higgs sector 
in Sec.~\ref{sec:HiggsSector}. 
Sec.~\ref{sec:summary} is devoted to the summary.

\section{grand gauge-Higgs unification}
\label{sec:gGHU}
In Ref.~\cite{gGHU}, a possibility to apply the Hosotani 
mechanism~\cite{hosotani} to the GUT breaking is revisited. 

The Hosotani mechanism is one of the mechanisms for gauge symmetry 
breaking. 
It works on a higher-dimensional gauge theory with the extra-dimensions 
compactified (to reproduce our four dimensions in the phenomenological 
applications). 
In view of the four-dimensional (4D) effective theory, the extra-dimensional 
components of the gauge fields behaves as scalar fields and may develop 
non-vanishing vacuum expectation values (VEVs) to break the gauge symmetry. 
In other words, 
the gauge field 
and Higgs fields are unified, and thus often called the gauge-Higgs 
unification (GHU)~\cite{GHU} especially when it is applied to the 
electroweak symmetry breaking.  
We name our case where it is applied to the GUT breaking as the 
grand GHU~\cite{gGHU}\footnote{%
This name is already used in Ref.~\cite{LimMaru} in a slightly different 
context which we carelessly overlooked.}.

One feature of this mechanism is that the order parameter 
is a continuous Wilson loop defined by exponentiating the gauge field. 
This quantity is an element of the gauge group instead of the algebra. 
Thus, it 
is constrained 
to be special, ${\rm det}W=1$, instead of the traceless condition and, 
for instance, 
the form $\left< W\right>={\rm diag.}(1,1,1,-1,-1)\equiv P_W$ 
in the $SU(5)$ case
is allowed. 
Since the entry $1$ corresponds to the vanishing exponent and thus to 
the trivial vacuum, the VEV is effectively written as 
$P_W\sim{\rm diag.}(0,0,0,v,v)$ which is a kind of those called 
{\it missing VEVs} that has vanishing eigenvalues. 
The missing VEVs can be used to {\it split} the masses of different components 
in a single multiplet, especially of the doublet and the triplet. 
Usually, the missing VEV can not be realized in the $SU(5)$ model 
due to the traceless condition 
and larger GUT symmetry, such as $SO(10)$, is considered~\cite{DW}. 
Although the naively expected pattern of the mass splitting is opposite to 
the desired one, the VEV $P_W$ results in the desired one for anti-periodic 
bulk fields with vanishing mass terms~\cite{gGHU-DTS}. 
Thus, when the SUSY version of this scenario is considered to protect the mass 
terms from the quantum corrections, the DT splitting problem can be solved, 
assuming the VEV $P_W$ is obtained. 

In this way, the possibility of the GUT breaking via the Hosotani mechanism 
is interesting enough to be seriously examined. 
Once we start to try to construct a model on this line, however, 
we immediately face a difficulty. 
Since the higher-dimensional theories are essentially vector-like in view 
of the 4D theory, the extra-dimensions should be compactified 
on a so-called orbifold. 
The orbifold boundary conditions (BCs) that project out one chirality, however, 
tend to project out also the desired massless adjoint scalar fields. 
%
%
In Ref.~\cite{gGHU}, it is shown that the so-called diagonal embedding 
method~\cite{DiagonalEmbedding} invented in context of the string theory 
can be applied also in field-theoretical setups, as demonstrated below. 
An advantage of the field-theoretical setup is that it is much easier to 
calculate the quantum corrections to the scalar potential which is calculable, 
{\it i.e.} finite~\cite{finiteness}, and determines the position of the vacuum. 
In the simplest case, an 5D $SU(5)$ model compactified on an $S^1/\Z2$ 
orbifold, the VEV $P_W$ can be realized as a minimum for appropriate matter 
contents~\cite{gGHU}, with no need of fine-tuning as a $\Z2$ symmetry is 
recovered on this vacuum~\cite{gGHU-DTS}. 

In order to apply the diagonal embedding method to the above simplest case, 
we impose two copies of the $SU(5)$ gauge symmetry with a $\Z2$ 
symmetry that exchanges the two $SU(5)$, and set the BCs around the 
endpoints of the $S^1/\Z2$ as the two gauge fields are exchanged.
It is straightforward to see that, with these BCs, the gauge symmetry remaining 
unbroken in the 4D effective theory is the diagonal part of the 
$SU(5)\times SU(5)$ 
(or our GUT symmetry is {\it embedded} into the {\it diagonal} part) and an 
adjoint scalar field is actually realized. 
These become apparent when the 5th dimension is put on a lattice 
which is very similar to the one of $S^1$ model, as shown in 
FIG. 1 in Ref.~\cite{gGHU}.
%
%
A difference from the $S^1$ model is that there is a brane at each endpoint 
of the $S^1/\Z2$ on which chiral fermions can be put.
Namely, our setup can be seen as a way to {\it introduce} chiral fermions 
in $S^1$ models. 
An interesting feature is that the bulk fields are similar to those in $S^1$ 
models and couple to the VEV $P_W$, and thus the zero modes appear 
as vector-like $SU(5)$ incomplete multiplets as the two Higgs doublets in the 
minimal SUSY SM (MSSM). 
On the other hand, the boundary fields are essentially 4D fields and do not 
couple to the VEV, and thus appear as chiral $SU(5)$ complete multiplets 
as the MSSM matter multiplets. 



\section{low energy effective theory}
\label{sec:LEET}
Now, we examine the properties of the low energy effective theory of the SGGHU. 
By definition, in this scenario, there is an adjoint scalar field that originates 
from the gauge field, which is massless at the tree level. 
Its mass is generated by the quantum corrections which vanish in the limit 
of the exact SUSY. 
This means that the loop-induced mass is at most of order of the SUSY-breaking 
scale.
The masses of its SUSY partners can differ from its mass again at most by the
SUSY-breaking scale. 
As a result, adjoint chiral multiplets, in concrete an color octet, a weak 
triplet and a singlet all with vanishing hypercharge, with mass around the 
SUSY-breaking scale generally exist in the SGGHU~\cite{gGHU-DTS}. 
Existence of these adjoints can be tested in future collider experiments 
if the SUSY-breaking scale lies around the TeV scale. 

An immediate consequence of the existence is, however, that the success of the 
GCU is dismissed, although the unification can be recovered 
by introducing additional bulk matter fields whose zero modes consist of $SU(5)$ 
incomplete multiplets.
An example we consider is that the additional zero modes are two vector-like pairs 
of $(\bar{L},L)$ $(({\bf1},{\bf2})_{-1/2})$, one of $(\bar{U},U)$ $((\bar
{\bf3},{\bf1})_{-2/3})$ and one of $(\bar{E},E)$ 
$(({\bf1},{\bf1})_{1})$~\cite{gGHU-DTS}.
With these matter content, the three gauge couplings meet with each others 
at the GUT scale with a unified gauge coupling $\alpha_G\sim 0.3$. 
This value is still perturbative, but is rather strong. 
In particular, the $SU(3)$ is not asymptotic free and the quantum corrections 
to the colored particles are enhanced, and for instance the $\mu$ parameter of 
the octet becomes 70 times larger than the one of the triplet at the low 
energy~\cite{gGHU-pheno}. 

\section{Higgs sector}
\label{sec:HiggsSector}
Since the colored particles tend to be too heavy to study at the LHC 
as mentioned above, below we focus on the colorless adjoints, {\it i.e.} the 
triplet $\Delta$ and the singlet $S$ which enlarge the Higgs sector. 
Now, this sector is composed of the MSSM two Higgs doublets $H_u$ and $H_d$, 
$\Delta$ and $S$.  
The superpotential is given by
\begin{eqnarray} \label{eq:WHiggs}
  W=\mu H_u \cdot H_d+\mu_\Delta{\rm tr}(\Delta^2)+\frac{\mu_S}2S^2
  +\lambda_\Delta H_u \cdot \Delta H_d + \lambda_SSH_u \cdot H_d.
\end{eqnarray}
Notice that there are no self-interactions among $S$ and $\Delta$, although 
such couplings are not forbidden by the symmetry of the effective theory, 
as $S$ and $\Delta$ originate from the gauge supermultiplet.  

The two new couplings $\lambda_\Delta$ and $\lambda_S$ 
push up the SM-like Higgs boson mass via the $F$-term contributions as in the 
next-to-MSSM (NMSSM)~\cite{NMSSM} and 
cause mixing between the MSSM doublets and the adjoints which results in 
modification of the SM-like Higgs coupling constants. 
Interestingly, these couplings are related with the unified gauge coupling 
$g_{\rm GUT}$ 
as $\lambda_\Delta =2\sqrt{5/3}\lambda_S (\equiv\lambda_S')=g_{\rm GUT}$ 
at the GUT scale.  
Thus, this model is very predictive, up to the SUSY-breaking parameters.  
To be more concrete, for the above example of the additional chiral matter
multiplets to recover the GCU, we obtain $\lambda_\Delta=1.1$ and 
$\lambda_S=0.25$ at the TeV scale~\cite{gGHU-pheno}.

\begin{figure}[t]
\includegraphics[width=45mm]{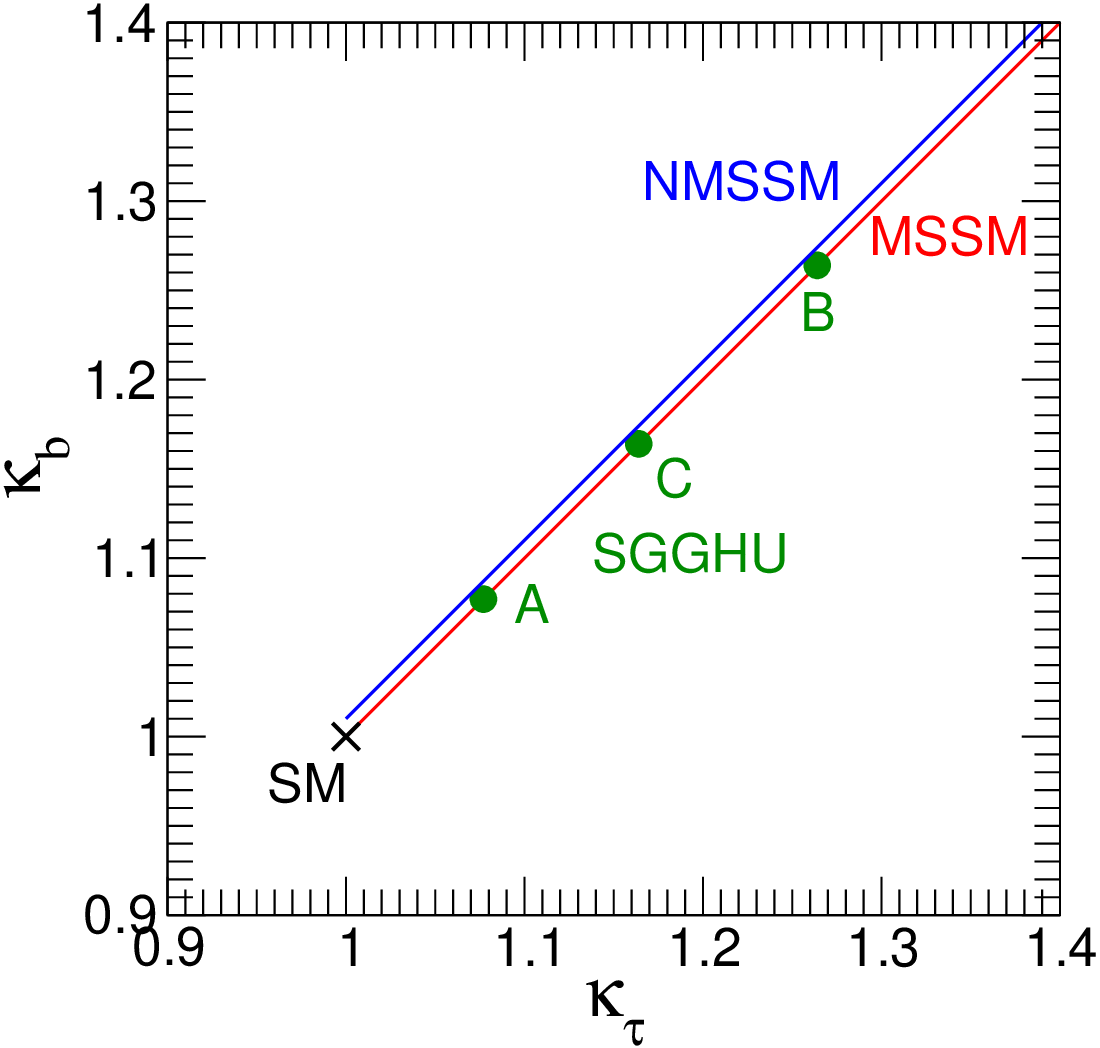} \qquad
\includegraphics[width=45mm]{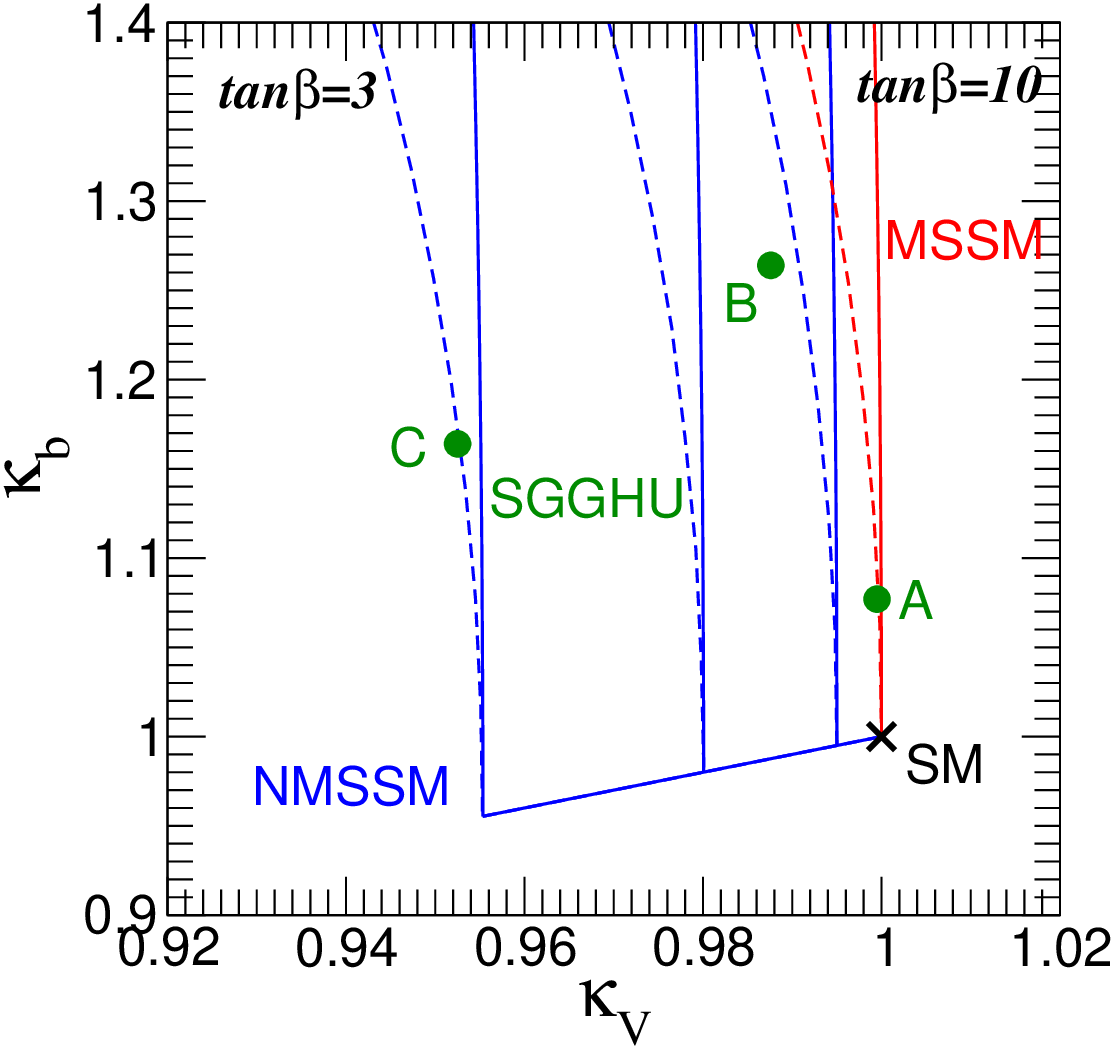} \qquad
\includegraphics[width=45mm]{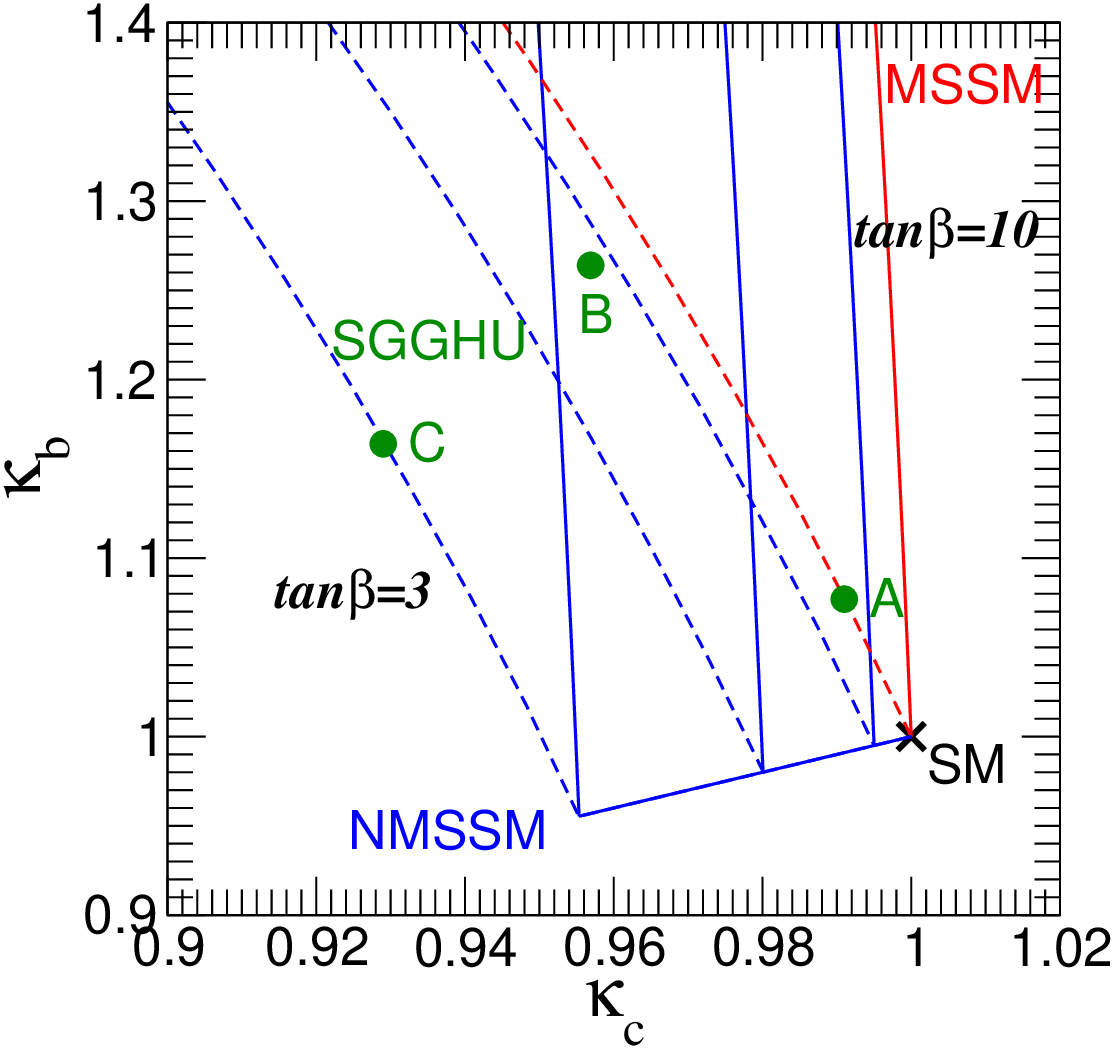}
\caption{\footnotesize The deviations in the SM-like Higgs couplings to a 
  SM field $X$ from the SM predictions $\kappa_X$ are plotted.
  The predictions of the three benchmark
  points (A), (B) and (C) in the SGGHU are shown with green blobs.
  The MSSM and NMSSM predictions are shown with red and blue lines,
  respectively, for $\tan \beta=10$ (thick line) and $\tan \beta=3$ (dashed). 
  Three lines for the NMSSM predictions indicate mixings between the
  SM-like and singlet-like Higgs bosons of 10\%, 20\% and 30\% from
  the right to the left.
  For the purpose of illustration, the NMSSM line is
  slightly displaced from $\kappa_\tau = \kappa_b$ in the left figure. }
\label{Fig:HiggsCouplings}
\end{figure}

As for the SUSY-breaking parameters, since the unified gauge coupling is 
strong, the unified gaugino mass must be large, say 3 TeV.  
As a result, soft masses of the colorless fields at the TeV scale are  
typically $1$-$2$ TeV.  
In addition, as in the MSSM, the soft term of $H_u$ receives a large 
contribution via the large top Yukawa interaction.  
Therefore, some tuning is unfortunately needed to realize electroweak 
symmetry breaking, and the higgsino mass parameter
$\mu$ and the CP-odd Higgs boson mass $m_A$ also tend to be $1$-$4$ TeV.  
In order to realize scenarios where some of the extra Higgs
boson masses are of the order of ${\cal O}(100)$ GeV, further tuning
is required among the input parameters.
Therefore, we will show some benchmark points
that reproduce the mass of the SM-like Higgs boson, instead of
scanning the parameter space.  
We focus on the following three different cases:
\begin{itemize}
\item[(A)] All the Higgs bosons other than the SM-like Higgs boson are heavy.
\item[(B)] The new Higgs bosons other than the MSSM-like Higgs bosons are heavy.
\item[(C)] The new Higgs bosons affect the SM-like Higgs boson couplings.
\end{itemize}
Bearing the fact that there are a few GeV uncertainties in the
numerical computation of the SM-like Higgs boson mass, we take the
range of $122~{\rm GeV} < m_h < 129~{\rm GeV}$ as its allowed region.
Values of
parameters of the TeV-scale effective theory are obtained after RG
running and shown in Table \ref{tab:benchmark-TeV}.  
%
\begin{table}[t]
  \caption{\footnotesize Examples of parameters of the TeV-scale effective theory
    obtained after RG running. For all the cases, $\tan\beta=3$, the gaugino masses 
    are $(M_1,M_2,M_3)=(194, 388, 1390)~{\rm GeV}$ and the adjoint $\mu$ terms are 
    $(\mu_\Delta,\mu_S)=(-252, -85.8)~{\rm GeV}$.
    }
  \begin{center}
    {\footnotesize
%
  \begin{tabular}{|c||c|c|c|c|c|c|c|c||c|}
    \hline
    Case &
    $\mu$ &
    $B \mu$ & 
    $\widetilde{m}_{\Delta}$ & 
    $\widetilde{m}_{S}$ &
    $\lambda_\Delta A_\Delta$ &
    $\lambda'_S A_S$ &
    $B_\Delta \mu_\Delta$ &
    $B_S \mu_S$ &
    $m_h$
    \\ \hline  \hline
    (A) &
    $177~{\rm GeV}$ &
    $42100~{\rm GeV}^2$ &
    $585~{\rm GeV}$ & 
    $195~{\rm GeV}$ &
    $284~{\rm GeV}$ &
    $446~{\rm GeV}$ &
    $288000~{\rm GeV}^2$ &
    $-5750~{\rm GeV}^2$ &
    $123~{\rm GeV}$
    \\ \hline 
    (B) &
    $177~{\rm GeV}$ &
    $40800~{\rm GeV}^2$ &
    $784~{\rm GeV}$ & 
    $612~{\rm GeV}$ &
    $1340~{\rm GeV}$ &
    $1110~{\rm GeV}$ &
    $30700~{\rm GeV}^2$ &
    $-110000~{\rm GeV}^2$ &
    $123~{\rm GeV}$
    \\ \hline 
    (C) &
    $175~{\rm GeV}$ &
    $41800~{\rm GeV}^2$ &
    $548~{\rm GeV}$ & 
    $216~{\rm GeV}$ &
    $284~{\rm GeV}$ &
    $446~{\rm GeV}$ &
    $207000~{\rm GeV}^2$ &
    $-33600~{\rm GeV}^2$ &
    $122~{\rm GeV}$
    \\ \hline 
  \end{tabular}
}
  \end{center}
  \label{tab:benchmark-TeV}
\end{table}
For these benchmark points, we calculate the deviation parameters
$\kappa_X = \frac{g_{hXX}}{g_{hXX}|_{\rm SM}}$
where $X$ denotes SM particles, as shown in Figure~\ref{Fig:HiggsCouplings}. 
There the predictions of the three benchmark
points (A), (B) and (C) in the SGGHU, in addition to those of the 
MSSM and NMSSM for comparison, are shown. 

Since the triplet mass has to be rather heavy to make its VEV of the neutral
component smaller than $10~{\rm GeV}$ in order to satisfy the rho parameter
constraint,  
the predictions of the SGGHU are not different from those of the NMSSM. 
In order to discriminate the SGGHU from the NMSSM, the mass difference 
among the MSSM-like additional Higgs bosons is useful.
The MSSM-like charged Higgs boson mass
$m_{H^\pm}^{}$ is given by
\begin{equation} 
  m_{H^\pm}^2
  = m_{H^\pm}^2|_{\rm MSSM}^{} (1 +\delta_{H^\pm}^{})^2 
  \simeq m_A^2 +m_W^2 +\frac{1}{8}\lambda_\Delta^2 v^2
  -\frac{1}{2}\lambda_S^2v^2\, ,
\label{Eq:mHpm}
\end{equation}
where $\delta_{H^\pm}$ is the deviation in $m_{H^\pm}^{}$ from the
MSSM and $m_A^{}$ is the MSSM-like CP-odd Higgs boson mass.  The sign
of the singlet contribution is opposite to the triplet one due to the
group theory.
For the couplings 
$\lambda_\Delta = 1.1$ and $\lambda_S = 0.25$ 
obtained above, the triplet contribution dominates and the deviation parameter
$\delta_{H^\pm}$ change the sign compared with the NMSSM case~\cite{gGHU-pheno}. 
As shown in FIG. 6 in Ref.~\cite{gGHU-pheno}, this parameter can be 
a few percent for $m_A\lesssim500~{\rm GeV}$. 
Since the charged Higgs boson mass can be determined with an accuracy
of a few percent at the LHC given such small masses \cite{HiggsWG}, we
can test our model.
%

When the masses of the adjoint-like scalars are
below $500~{\rm GeV}$, the International Linear Collider (ILC) has capability 
to directly
produce these new particles.  For example, the benchmark point (C)
gives the mass of the lighter
triplet-like scalar $\Delta^\pm$ is less than 
$500~{\rm GeV}$~\cite{gGHU-pheno},
and we can probe $\Delta^\pm$ using the channel $e^+ e^- \rightarrow
\Delta^+ \Delta^- \rightarrow tb \bar{t} \bar{b}$, which proceeds via
the mixing between the MSSM-like and triplet-like charged Higgs bosons.

\section{summary}
\label{sec:summary}
In the gGHU, the adjoint scalar field is realized in the 4D effective 
theory via the diagonal embedding method. 
Its VEV is determined by the calculable (finite) loop-induce scalar potential, 
and the VEV $P_W$ can be obtained without fine tuning in $SU(5)$ model with 
appropriate matter content. 
Assuming this {\it missing} VEV, the DT splitting problem can be solved in the 
SUSY version, and thus the SGGHU is theoretically well-motivated. 
In order to suppress the nucleon decay via the exchange of the gauge fields,
the scale of the VEV should be at least the usual GUT scale,  
beyond the reach of direct search at any feasible experiments. 
The SGGHU, however,  generally derives a testable prediction 
that adjoint chiral superfields with masses around TeV scale exist, 
and is attractive also from a viewpoint of the phenomenology. 

Since the colored particles become rather heavy in the SGGHU, we study its
Higgs sector. 
We show that the the SM-like Higgs mass is enhanced by the $F$-term 
contributions and that the deviations of the couplings from the SM
predictions are a few percent, which is a good target of future
electron-positron colliders, when the adjoint masses are below 
$1~{\rm TeV}$. 
The mass gap between the MSSM-like charged
Higgs boson and CP-odd Higgs boson differs from that of
the MSSM by a few percent, 
which 
is within the scope of
the LHC, when their masses are
below $500~{\rm GeV}$. 
Since the direction of the deviation is opposite to that in the NMSSM, 
the SGGHU can be discriminated from the other models by these measurements. 


\end{document}